# Benzophenone Semicarbazones as Potential α-Glucosidase and Prolyl Endopeptidase Inhibitor: *In-vitro* free radical scavenging, enzyme inhibition, mechanistic, and molecular docking studies

Qurat-ul-Ain*[a] Sidra Rafi[c], Khairullah[c], Saeedullah[c], Arshia Arshia[c] Reaz Uddin[a], Atia-ul-Wahab[a], Khalid Mohammed Khan[c], and M. Iqbal Choudhary*[c, d]

[a] Dr. Panjwani Center for Molecular Medicine and Drug Research, InternationalCenter for Chemical and Biological Sciences, University of Karachi, Karachi-75270, Pakistan

[c] H. E. J. Research Institute of Chemistry, International Center for Chemical and Biological Sciences, University of Karachi, Karachi-75270, Pakistan

[d] Department of Biochemistry, Faculty of Science, King Abdulaziz University, Jeddah 21589, Saudi Arabia

**ABSTRACT:** α-glucosidase and prolylendopeptidase has altered expression and activity patterns in neurological disease, type 2diabetes respectively and several cancers. Here we screened a series **1-29** benzophenone semicarbazone derivatives for in vitro free radical scavenging, α-glucosidase and prolylendopeptidase inhibition activities. Seven derivatives were identified as potential free radical scavengers, **14** as α-glucosidase, and **9** as prolylendopeptidase inhibitors. Kinetic studies on the most promising inhibitors were performed. Compounds **23, 27, 25** and **28** were found as inhibitor of α-glucosidase, while compound **26** inhibited both prolylendopeptidase and α-glucosidase. The binding modes and binding free energy of the multi targeted inhibitor **26** were predicted by molecular docking studies. The top-ranked conformation of compound **26** was found to have -8.25 kcal/mol estimated binding free energy with $\alpha$-glucosidase compared to the binding affinity of -7.89 kcal/mol of its standard acarbose. In addition, prolyl endopeptidase with compound **26** contains -7.17 kcal/mol, while the prolinal has the binding free energy of -7.63 kcal/mol. These results provide insights on prolylendopeptidase and α-glucosidase inhibition of compound **26** for further development as therapeutic agents for neoplastic, neurological, and endocrine disorders.

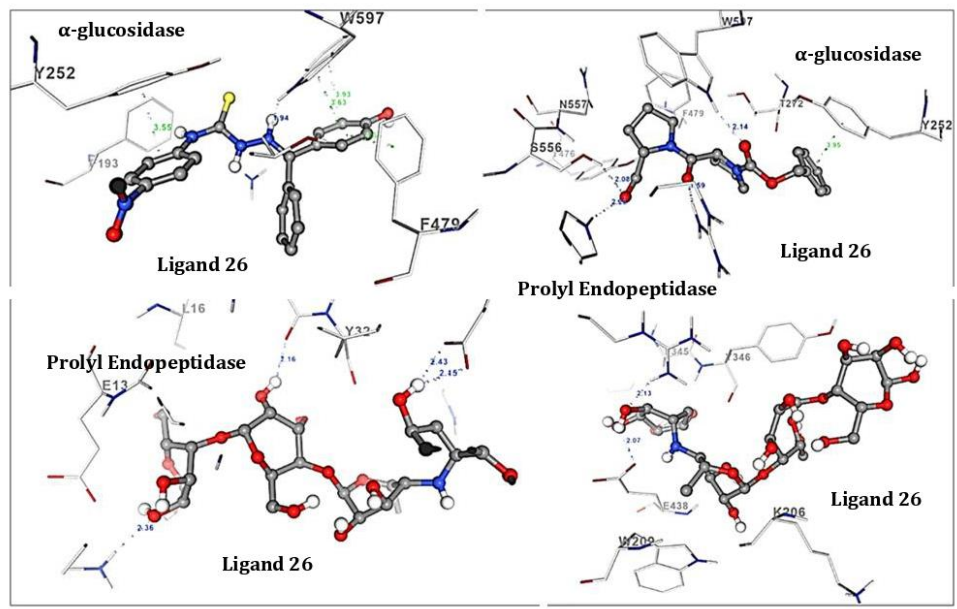

## INTRODUCTION:

α-Glucosidases (EC 3.2.1.20), as a carbohydrate digestive enzyme, controls the production of glucose by catalyzing the breaking down long-chain polysaccharides into monosaccharide.[1-2] High concentration of monosaccharides and polysaccharides in the body resulting from the action of α-glucosidases lead to hyperglycemia that is considered as one of the hall mark of type 2 diabetes (T2DM).[3-6] The inhibition of α-glucosidases activity is one of the prominent approach to modulate the onset of postprandial hyperglycemia, thereby rendering it an ideal target for the management of T2DM.[7-9] Currently, T2DM is a major healthcare concern worldwide and the most common endocrine disorder in the United States, with over 10% of Americans struggling with either type 1 or type 2 diabetes.[10,11] Although a number of antidiabetic agents are currently in clinical use but due to some side effects of the available α-Glucosidase inhibitors such as flatulence, diarrhea, and abdominal discomfort, a need of new safe α-glucosidase inhibitors exists.[12-16]

Cancer remains a high-risk global health problem, during 2022 more than 1.9 million new cancer cases were diagnosed only in the US while 609,360 deaths were reported.[17,18] Approximately, 90% of cancer patients died due to its metastasis not from the primary tumor.[19, 20] Carbohydrate residues on cell-surface glycoconjugates are suggested to contribute to cancer metastasis together with various other factors by modifying malignant phenotypes.[21,22] α-glucosidase inhibitors were shown to inhibit cancer metastasis by interfering biosynthesis and oligosaccharide structure on cancer cell surface.[22] An α-glucosidase inhibitor castanospermine inhibit metastasis by disturbing the carbohydrate structure on the surface of neoplastic cells.[23, 24] Currently available anticancer drugs exhibit various side effects by working on an unspecified target, and marred by drug resistance.[24] Therefore, discovering small molecules that exhibit potent and selective anticancer activities is still a goal and challenge in medicinal chemistry research.

Prolyl endopeptidase (PEP; EC 3. 4. 21. 26) is a serine protease, cleaves proline containing neuropeptides. PEP hydrolyzed peptide bond at the C-terminal of internal proline residue. [25] Over activity and expression of this ubiquitous enzyme observed throughout normal human and other mammalian tissues.[26] This enzyme involved in differentiation, development, and proliferation processes of several tissues and the altered patterns of the expression and catalytic function of peptidases may contribute to tumor progression, therefore PEP is a potential target for cancer therapy.[27] PEP expression increases in the brain with age has been implicated AD that is a progressive neurodegenerative disorder characterized the age-related dementia, by memory loss and cognitive decline.[28, 29] It is estimated that currently about 50 million people worldwide suffer from AD or other dementia, and the number is expected to reach 152 million by 2050. [30] To date, there is still no effective treatment for this irreversible and devastating disease. It has been recognized that inhibitors of PEP may alter the activity of neural network and adjust neuropeptides levels to normal in AD.[31-33] PEP inhibitors have been proposed as potential therapeutic agents for some of these disorders. Substantial epidemiological evidence indicates an increased risk of developing AD in patients with type 2 diabetes.[34] Researchers have identified physiological factors that are associated with both T2DM and AD, including neuro degeneration, β-amyloid (Aβ) deposition, glycogen synthesis kinase3, tau-protein phosphorylation, oxidative stress and most important altered insulin signaling[35]

Free radicals that contain oxygen reactive species (ROS) such as superoxide radical anion, hydrogen peroxide, and hydroxyl radical ions are highly reactive molecular species their excessive level cause damage the integrity of various biomolecules including lipids, proteins, and DNA via producing oxidative stress (OS).[36, 37] The excess levels of free radicals result due to enhanced ROS production and/or decreased ROS scavenging ability. In fact, the pathology of several diseases including several type of cancers, diabetes mellitus, neurodegenerative diseases, rheumatoid arthritis, cataracts, cardiovascular and respiratory diseases are greatly influenced by oxidative stress.[38] Several reports highlight a close connection between insulin resistance and defects in energy metabolism driven by OS. [39] AD patients show reduced brain insulin receptor (IR) sensitivity Therefore, antioxidants/free radical scavengers could be useful to control ROS associated tissue damaged through the scavenging and/or downregulating of ROS generation. Therefore, discovery of a novel molecule exhibiting both prolylendopeptidase and α-glucosidase together with potential of scavenging free radicals could be act as a lead agent to treat shared disease.

Benzophenone semi carbazones form an important class of biologically active compounds as this skeleton has been previously reported to confer various biological activitiesto the compounds, including anti-cancer, ant-amicrobial, anti-inflammatory, anti-human immunodeficiency virus, urease inhibition, anti-plasmodial, anti-glycation, and anti-oxidant,[40] activities. Surprisingly none of the benzophenone semicarbazones was tested previously for free radical scavenging, as well as α-glucosidase, and prolylendopeptidase PEP inhibition properties. Therefore, in the continuation of our work, on discovery of potent biological molecules asdrug leads, we screened newly synthesized substituted benzophenone semicarbazone (BS) derivatives for free radical scavenging, and α-glucosidase and prolylendopeptidase PEP inhibitory activities. We also studied binding modes of promising compound **26** and calculated binding free energies. Considering the good agreement of compound **26** with both the enzymes and preferred antiradical activities, we suggest compound **26** for further development as a novel drug that could be used to inhibit α-glucosidase, prolylendopeptidase enzymes and maintain intracellular oxidative stress for treatment of neoplastic, neurological, and endocrine disorders.

## ■ EXPERIMENTAL PROCEDURES

**Chemistry.** All benzophenone semicarbazone derivatives were obtained from the Drug Bank of the Dr.Panjwani Center for Molecular Medicine and Drug Research, ICCBS, University of Karachi, Pakistan.

**Source of Enzymes and Chemicals.** α-Glucosidase and *Flavobacterium* prolyl endopeptidase (PREP) were purchased from Sigma Aldrich (USA). Standard compounds, *i.e.,* acarbose and bacitracin, were obtained from Sigma Aldrich (USA), and butylated hydroxytoluene was purchased from Sigma Aldrich (Germany). Monobasic sodium and dibasic sodium phosphate buffers were acquired from Sigma Aldrich (USA). Solvent 1, 4-dioxane was obtained from Fisher Scientific (Germany), reagent grade ethanol and dimethyl sulfoxide (DMSO)) was purchased from Sigma Aldrich (USA). Substrates: *p*-nitrophenyl-α-D-glucopyranoside and Z-Gly-Pro-pNA (substrate) were obtained from Sigma Aldrich (Germany).

**Free Radical Scavenging Assay.** Free radical scavenging activity of benzophenone semicarbazone derivatives **1-29** were measured using DPPH radical by the method previously described.[41] Samples were dissolved in DMSO to obtain 0.5 mM, and further diluted to obtain different concentrations. 5 $\mu$L sample of each concentration was transferred to 96-well plate in triplet, and initial read was recorded at 517 nm. 0.3 mM solution of DPPH was prepared freshly in ethanol, and 95 $\mu$L of was added in each of the 96 wells. Finally, an absorbance was recorded at 517 nm. An equal concentration of butylated hydroxytoluene was tested as positive control and the absorbance of the equal volume of DMSO was recorded as negative control. Following formula was used to calculate the radical scavenging activities:

% Radical scavenging activity of DPPH = [$A_0$-$A_1$/$A_0$] ×100
Where:
$A_0$: The absorbance of all reagents without the tested compounds. $A_1$:
The absorbance in the presence of test compounds.

**In *Vitro* α-Glucosidase Inhibition Assay.** α-Glucosidase inhibition was measured spectrophotometrically using methods described by Adisakwattana *et. al.*[42] Different concentrations of test compounds were incubated with 1 μ/mL of enzyme, dissolved in pH 6.8 PBS buffer, for 15 min at 37 °C, then 0.7 mM of the *p*-nitrophenyl-α-D-glucopyranoside as substrate was added, reaction was allowed to proceed in multiplate reader upto 31 min, and hydrolysis of substrate by enzyme was recorded at 400 nm. Kinetics studies were carried out by adding different concentrations 0.1, 0.2, 0.4, and 0.8 mM. DMSO 7% and acarbose were used as blank and positive control respectively.

***In vitro* Prolyl Endopeptidase (PREP) Inhibition Assay.** Milk and Orlowski method with slight modification was employed to evaluate PEP inhibitory activity.[43] In each well of a 96-well plate, 200 μL of total reaction volume (20 μL of PEP enzyme dissolved in 140 μL of 50 μ*M* phosphate buffer at pH 7.0, and 20 μL of 0.5 mM test compound) was incubated at 30 °C. As a substrate 0.4 μ*M* of 20 μL of Z-Gly-Pro-pNA, dissolved in 40% 1, 4 dioxane, was added in the reaction mixtures after 10 minutes of incubation. The components were allowed to react for 30 min then at 410 nm a absorbance was measured by using Multiskan GO (Thermo Fisher Scientific OyRatastie 2, P.O. Box 100 FI-01621, Finland). The $IC_{50}$ values were calculated using EZ-FIT enzyme kinetics software (Perrella Scientific, Inc., Amherst, NH 03031, USA). For mechanistic studies, 250 to 32 μ*M* of test compounds was incubated with 0.02U/200 μL PEP enzyme for 10 min at 30 °C. Different concentrations (0.20 to 0.50 mM) of Z-Gly-Pro-pNA were then added to initiate reaction. ELISA plate reader (SpectraMax 384, Molecular Devices, USA) was used to monitor the cleavage of Z-Gly-Pro-pNA at 410 nm for 30 min.

**Compound Preparation.** The Chemdraw was used to draw the two-dimensional structure of the most promising compound **26**. 3D atomic coordinates were generated *via* distance geometry using molecular connectivity through Babel, and the multi-objective genetic algorithm was used to ensemble conformer.[44] Energy minimization was performed using FROG2.[45] to remove clashes among atoms of the ligand and develop a reasonable starting geometry. Fifty conformers were generated with 100 maximal number of Monte-Carlo steps, and the energy window was set to 100. The reference compounds, acarbose and prolinal, for enzyme α-glucosidase and prolyl endopeptidase (PEP), respectively, were downloaded from Drug Bank.[46]

**Protein Modelling.** The web-based tools were employed for model refinement, evaluation, tertiary structure, and binding pockets predictions.

**Swiss-Model.** The 3D structure of prolyl endopeptidase (PEP) is not available in any database; therefore, we used a comparative modeling approach to predict the tertiary structure of PEP. The UniProt database was used to retrieve the FASTA format sequence having identifier number P27028.[47] PEP is a full-length protein sequence containing 705 amino acid residues.

**Quality Assessment.** The stereochemical quality of the predicted protein structure was analyzed *via* residue-by-residue geometry by producing Ramachandran plots developed using PROCHECK.[48]

**Molecular Docking.** For binding mode prediction of compound **26**, and substrate with enzymes, the docking studies were performed using AutoDock Vina version 1.5.6, supported by the MGL tool. [49, 50] The predicted tertiary structure of prolyl endopeptidase (PEP) and α-glucosidase was docked with compound **26** and their standard inhibitors acarbose and

Z_Pro_Prolinal, respectively, to validate our computational results. The structure was centered in a grid box by setting the number of points in each dimension X, Y, and Z and 0.375 Å spacing between points. The genetic algorithm and Lamarckian genetic algorithm were used,[51] and the parameters were set to the number of conformations 150, while the population size was 500 with utmost figure of energy evaluations set to 25,00,000 and number of generations 27,000. The gene mutation rate was calculated as 0.02. The top-ranked pose was selected to explore the binding interactions with their binding receptors. For that Discovery studio (REF) was used, which gives 2D/3D visualization of ligand-receptor interactions.[52]

## ▪ RESULTS AND DISCUSSION

**Banzophenone Semicarbazones Derivatives Exhibit Variable Degree of Radical Scavenging Activities.** A library of newly synthesized benzophenone semicarbazone derivatives **1-29** was evalaed for is free radical scavenging activity (Table 3). Derivatives **16**, **23**, **25**, **26**, **27**, **28**, and **29** were found to scavenge DPPH free radical in a concentration dependent manner, as depicted in Figure 1. Compound **26** was found most potent radical scavenger of the series due to exhibiting 80.14% antioxidant activity against DPPH radicals compared to the butylated hydroxytoluene (standard). This standard in an *in vitro* assay found to display a significant antioxidant activity by scavenging 85.8% of DPPH radicals.

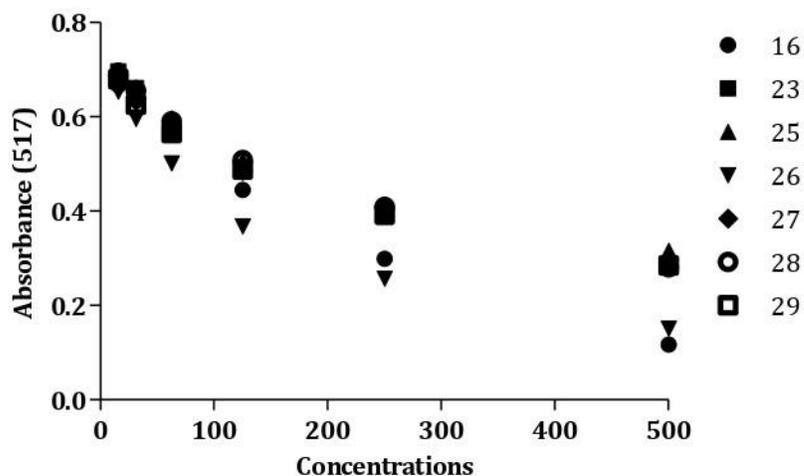

**Figure 1. Concentration Dependent Radical Scavenging Activity of Selected Benzophenone Semicarbazone Derivatives.** Graph plotted between absorbance *vs* different concentrations of selected benzophenone semicarbazone redical scavengers showed a linear relationship between radical scavenging capacity and concentrations of compounds.

**α-Glucosidase Inhibition Studies.** Derivatives **1–28** derivatives were evaluated for their capacity of α-glucosidase inhibition, and fourteen compounds **1**, **3**, **4**, **8**, **9**, **18**, **19**, and **20-28** were showed significant inhibition of the enzyme with $IC_{50}$ values between 27.69 to

42.29 μM. However, the enzyme was most significantly inhibited by the compounds **1**, **18**, **20**, **26** with $IC_{50}$ values of 26.76 ± 2.14, 24.62 ± 1.49, 29.43 ± 2.04, and 27.69 ± 2.04 μM, respectively, as compared to acarbose standard ($IC_{50}$ = 875.75 ± 2.08 μM) as presented in Table

**2**. Total 5 compounds were selected to study their modes of inhibition through mechanistic studies, as presented in Table-1.

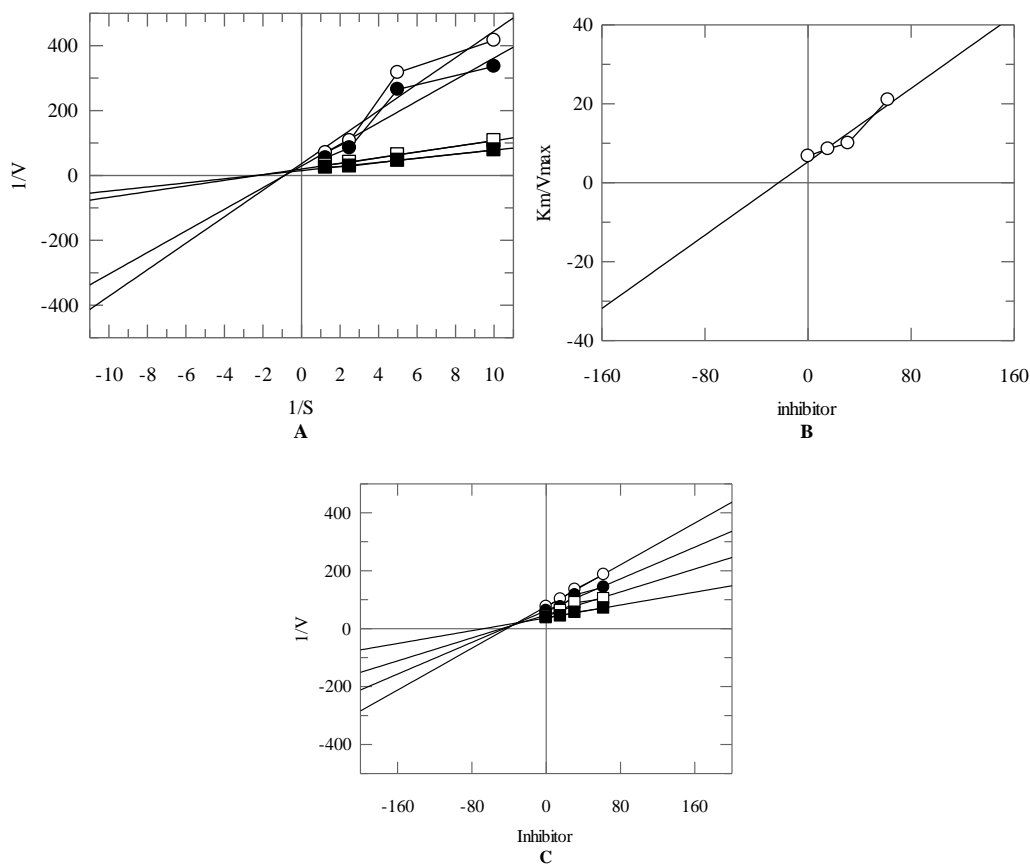

**Figure 2. Compound 26 mediated mixed inhibition of α-glucosidase.**
(A) Lineweaver-Burk plot of reciprocal of rate of reaction (V) *vs* reciprocal of substrate without (■), and with 62.00 $\mu M$ (○), 31.00 $\mu M$ (●), and 15.00 $\mu M$ (□) of compound **26**. (B) Secondary replot of Lineweaver-Burk plot between the slopes of each line on Lineweaver-Burk plot *vs* varying concentrations of compound **26**. (C) Dixon plot of reciprocal of rate of reaction (V) *vs* varying concentrations of compound **26**.

These compounds were found to induce inhibition of enzyme by different ways, among them compounds **23** and **27** have showed competitive-type of inhibition in which *Vmax* remains constant and *Km* value decreases, as deduced from the Lineweaver-Burk plot (Figure 1A, and Figure 3A of the supporting information).

**Table 1.** Kinetics of most active compound on α-glucosidase.

| Compounds No. | IC$_{50}$$^a$ ± SEM$^b$ (μM) | Ki* ± SEM$^b$ (μM) | Type of inhibition |
|---|---|---|---|
| 23 | 31.62 ± 1.46 | 26.27 ± 0.013 μM | Competitive |
| 25 | 38.61 ± 1.43 | 32.22 ± 0.008 μM | Mixed |
| 26 | 27.69 ± 2.04 | 22.73 ± 0.017 μM | Mixed |
| 27 | 35.24 ± 1.89 | 31.28 ± 0.013 μM | Competitive |
| 28 | 42.29 ± 1.76 | 37.26 ± 0.016 μM | Mixed |

Ki * = Concentration required to achieve half maximal inhibition was deduced from Dixon plot and presented as mean of five values, a IC$_{50}$ is the concentration required to achieved 50% inhibition , b SEM = Standard mean error of 3–5 experiments.

Compound **26** was found to inhibit α-glucosidase by mixed type on inhibition Figure 2A. However, the Lineweaver-Burk plot of compounds **25** and **28** showed that these compounds were also exhibited a mixed-type of inhibition (Figure 2A and Figure 4A of the supporting information), in such typeof inhibition, *Km* values some time increase or decrease, but *Vmax* always decreases. The concentrations of these inhibitors required to produce half maximum inhibition were in the rangeof 22.73–37.26 μM, presented as Ki values in (Table 1) that were calculated through secondaryre-plot of Lineweaver-Burk plot (Figure 2B and Figure 1-4B of the supporting information), and Dixon plot (Figure 2C and Figure 1-4C of the supporting information).

**Table-2. Results of in vitro biological evaluation of 1-29 benzophenone semicarbazone derivatives.**

| COMPOUND | STRUCTURE | A-GLUCOSIDASE INHIBITION IC$_{50}$± SEM$^A$ (μM) | DPPH RADICAL SCAVENGING ASSAYIC$_{50}$ ± SEM$^A$ (μM) | PROLYLENDO PEPTIDASE INHIBITION IC$_{50}$± SEM$^A$ (μM) |
|---|---|---|---|---|
| 1. | 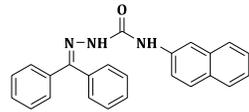 | 26.76 ± 2.14 | N/A | **N/A** |
| 2. | 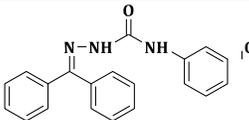 | Not available | N/A | **N/A** |
| 3. | 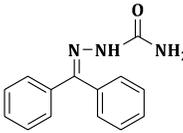 | N/A | N/A | **N/A** |
| 4. | 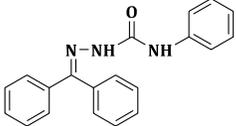 | 49.85 ± 1.82 | N/A | **N/A** |
| 5. | 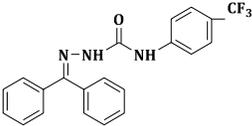 | N/A | N/A | **N/A** |
| 6. | 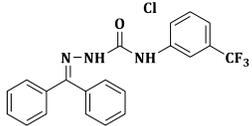 | N/A | N/A | **N/A** |
| 7. | 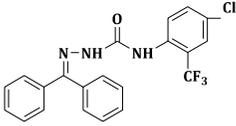 | N/A | N/A | **N/A** |
| 8. | 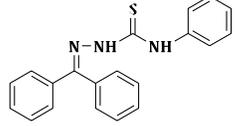 | 142.85 ±1.63 | N/A | **N/A** |
| 9. | 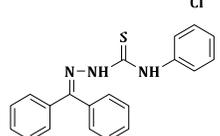 | 156.09 ± 1.24 | N/A | **368.5 ± 2.2** |
| 10. | 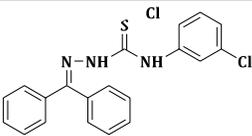 | N/A | N/A | **N/A** |
| 11. | 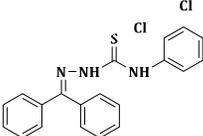 | N/A | N/A | **N/A** |
| 12. | 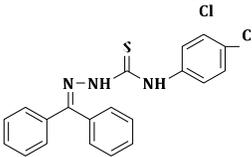 | N/A | N/A | **N/A** |
| 13. | 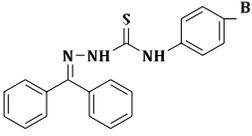 | N/A | N/A | **360.0 ± 2.1** |

| 14. | 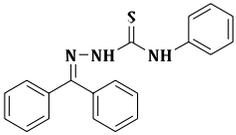 | N/A | N/A | **357.3 ± 1.8** |
|---|---|---|---|---|
| 15 | 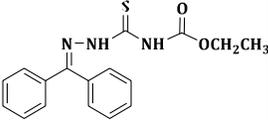 | N/A | N/A | **N/A** |
| 16 | 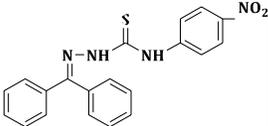 | N/A | 181.52± 2.2 | **N/A** |
| 17. | 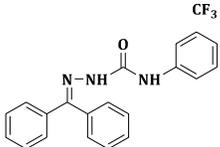 | N/A | N/A | **N/A** |
| 18. | 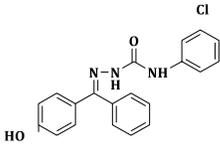 | 24.62 ± 1.49 | N/A | **400.8 ± 3.5** |
| 19. | 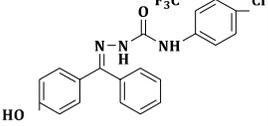 | 46.81 ± 1.78 | N/A | **N/A** |
| 20. | 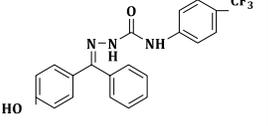 | 29.43 ± 2.04 | N/A | **412.4 ± 2.0** |
| 21. | 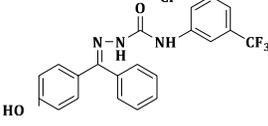 | N/A | N/A | **N/A** |
| 22. | 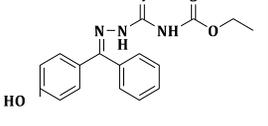 | N/A | N/A | **N/A2** |
| 23. | 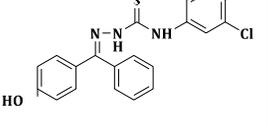 | 31.62 ± 1.46 | 351.8 ± 6.3 | **N/A2** |
| 24. | 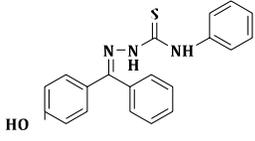 | 46.81 ± 2.31 | N/A | **N/A** |
| 25. | 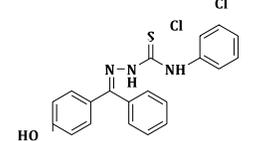 | 38.61 ± 1.43 | 351.83 ± 2.2 | **319.4 ± 2.2** |
| 26. | 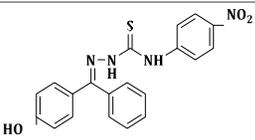 | 27.69 ± 2.04 | 123.3 ± 4.01 | **67.2 ± 1.3** |
| 27. | 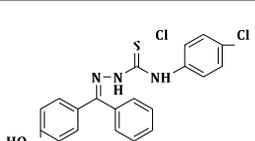 | 35.24 ± 1.89 | 309.192±5.3 | **N/A** |
| 28. | 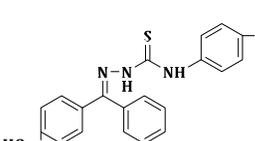 | 42.29 ± 1.76 | 307.2±1.8 | **440.0 ± 3.1** |
| 29. | 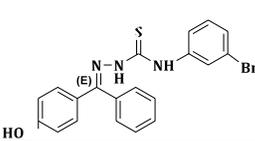 | 36.45 ± 1.91 | 282.2±2.8 | **423.5 ± 2.7** |
| STD | | **Acarbose 875.75 ± 2.08** | **BHT[b] 128.8 ± 2.1** | **Bacitracin 125 ± 1.5** |

a

SEM:standard error of mean, bBHT:butylated hydroxytoluene.

**Prolyl Endopeptidase Inhibition Studies.** Further studies were performed to evaluate the enzyme inhibitory activity of benzophenone semicarbazone derivatives **1–29** against Prolyl Endopeptidase (PEP). Among them 10 compounds showed significant inhibition of PEP (Table 2), in which compound **26** showed the most effective inhibitory activity (IC$_{50}$ =67.2 ± 1.3 $\mu$M), compared to standard inhibitor bacitracin (IC$_{50}$=125± 1.5 $\mu$M). Compounds **9, 13, 14, 18, 20, 25, 28,** and **29** showed a week inhibitory activity with IC$_{50}$ = 368.5 ± 2.2, 360.0 ± 2.1, 357.3 ± 1.8, 400.8 ± 3.5, 412.4 ± 2.0, 319.4 ± 2.2, 440.0 ± 3.1, and 423.5 ± 2.7$\mu$M, respectively. Compound **27** showed a significant inhibition of PEPwith IC$_{50}$ = 201.6 ± 2.6 $\mu$M, however remaining compounds were found to be inactive againstPEP (Table 2). Based on IC$_{50}$ values, compounds **9, 13, 25, 14, 20,** and **25, 26- 29**, which showedinhibitory potential against the PEP, mechanistic studies of most effective derivative of the series *i.e.,* compound **26** were further carried out. The mode of inhibition of compound **26** withits IC$_{50}$ and *Ki* value is given in Table-3.

As deduced from the Lineweaver-Burk plot (Figure 3A), compounds **26** decreased the *Vmax* values without affecting the *Km* values, hence, it is a non-competitive inhibitor. Compound **26** binds with the enzyme other than the active site, as inferred from 1/Vmaxapp values of every inhibitory concentration on the Y-axis at each intersection point of lines from this plot. Binding of compound **26** causes a change in the structure and shape of the enzyme, therefore, it was no longer able to bind with the enzyme correctly hence affected the rate of the reaction, catalyzed by the enzyme. The Dixon plots was used to determined potency of inhibition (*Ki* values) (Figure 3C.) that was further confirmed through secondary replot (Figure 3B).

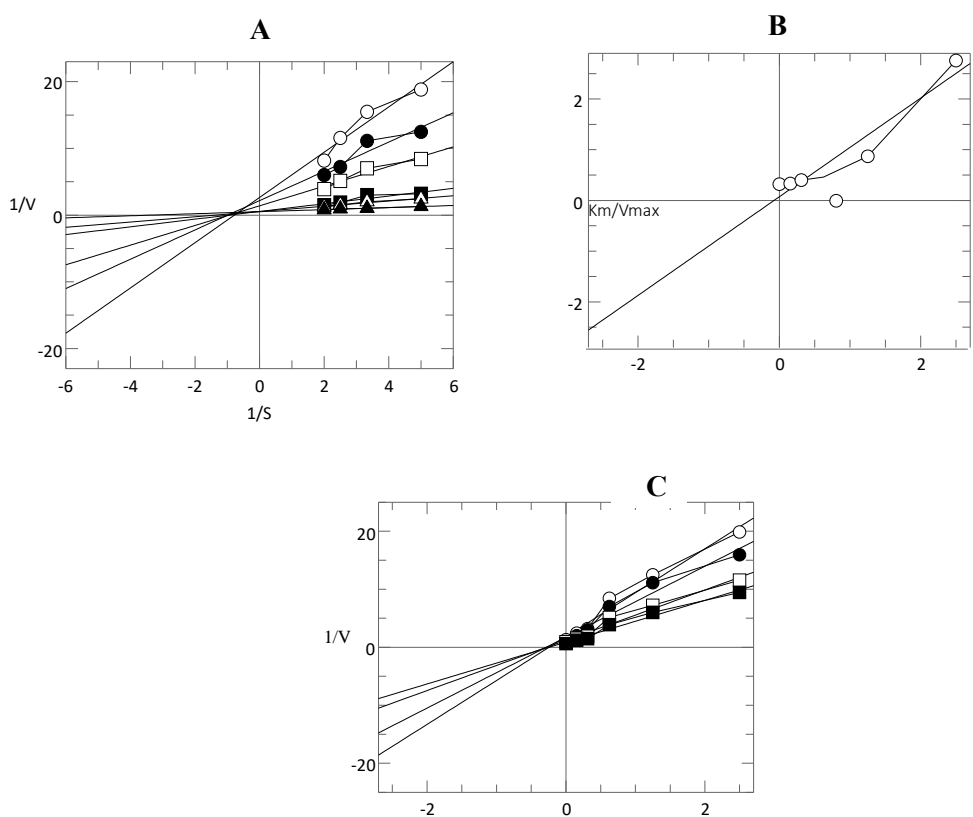

**Figure 3.** Non-Competitive Prolyl Endopeptidase (PEP) Inhibition by Compound **26**.

(A) Lineweaver-Burk plot, *i.e.* reciprocal rate of reaction *vs* reciprocal of substrate (Z-Gly-pNA) without (▲) and with 250.0 $\mu$M (○), 125.0 $\mu$M (●), 62.5 $\mu$M (□), 31.25 $\mu$M (■), 15.6 $\mu$M (Δ) of compound 26. (B). Secondary replot, *i.e.,* reciprocal of slope *vs* different concentrations of compound 26. (C) Dixon plot, *i.e.,* reciprocal of rate of reaction *vs* varying concentrations of compound 26.

**Table 3.** Kinetics of most active compounds **26** on Prolyl Endopeptidase.

| Compound | $IC_{50}{}^a \pm SEM^b$ ($\mu M$) | $Ki^* \pm SEM^b$ ($\mu M$) | Type of Inhibition |
|---|---|---|---|
| **26** | $67.2 \pm 1.3$ | $73.56 \pm 0.0055$ | non-competitive |

Ki * = Concentration required to achieved half maximal inhibition was deduced from Dixon plot and presented as mean of five values, a $IC_{50}$ is the concentration required to achieved 50% inhibition , b SEM = Standard mean error of 3–5 experiments.

**Computational Studies.** Finally, compound **26** was selected to study the binding mode prediction at an atomic level, as this compound exhibited remarkable free radical scavenging activity and potent inhibition for both prolyl endopeptidase (PEP) and α-glucosidase enzymes.

The 3D structure of *α*-glucosidase was obtained from PDB (PDB ID: 4J5T.[53] In contrast, the tertiary structure of PEP was not available in PDB or any other protein modeling database; therefore, we modeled PEP a through comparative modeling approach using the SWISS-MODEL protein homology-modeling server.[54]

The 3IVM is a template for the model protein, having a 56.66% of identity score. The QMEAN score is a composite scoring function, describing the significant geometrical aspects of protein structures having a -0.28 score, while the GMQE is a 0.80. The predicted structure is shown in Fig. 9 with their top ranked binding pocket using DoGSiteScore.[55] The binding pocket scores are given as a drug, and a simple score (Table S1 of the supporting information).

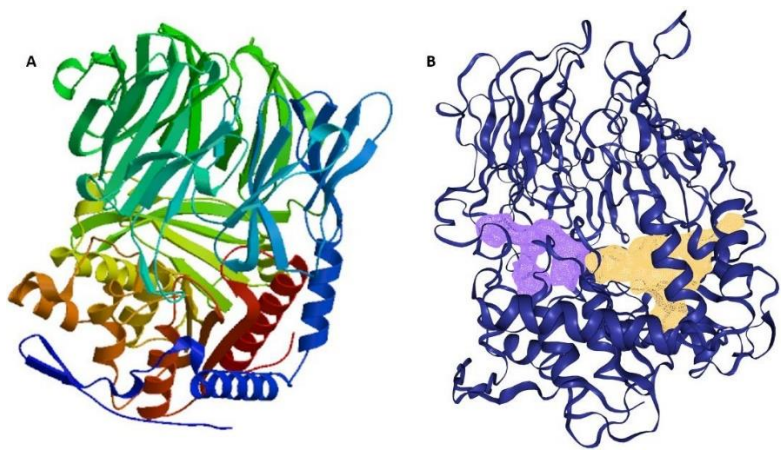

**Figure 4(A)** The modeled three-dimensional structure of prolyl endopeptidase (PEP) using Swiss-Model whereas **(B)** depicts the two potential binding sites of protein using DoGSiteScorer.

The quality of the predicted model was further validated by PROCHEK[56] using the Ramachandran plot Figure 4. The plot showed that 89.7% of residue occurred in the most favored region, whereas 9.3%, 0.2, and 0.8% lie in additional allowed, generous allowed, and disallowed regions.

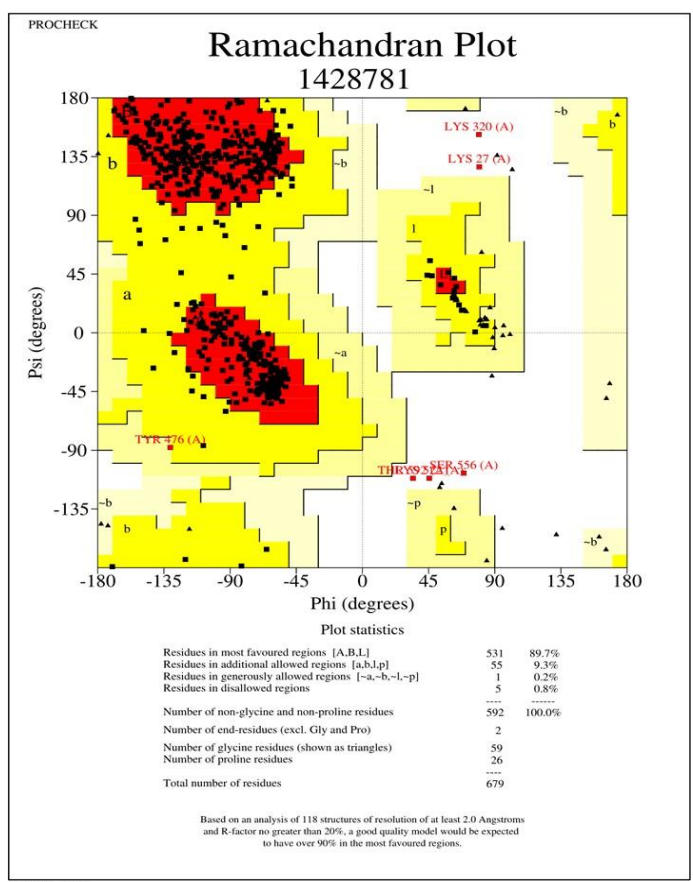

Figure 5: Ramachandran plot of model protein using PROCHECK.

☐ **Molecular Docking.** To gain some insight into ligand-protein binding modes and mechanisms, molecular docking studies were performed. Interestingly docking results of α-glucosidase and prolyl endopeptidase have shown good binding affinities with compound **26**.

☐ **Binding interactions of α-Glucosidase.** Docking study of compound **26** defined its interactions and specificity with the α-glucosidase as depicted in Figure 5. Various hydrogen bond acceptor and donor pairs with their suitable orientations and several hydrophobic interactions between the nonpolar active site residues and the compound were seen in the interaction pose. We observed the same binding site of ligands and standard, as shown in Figure 5A and S1 of the supporting information. The 2D interactions in Figure 5B (i) also showed a strong interaction of compound **26** with the active site residues of α-glucosidase. The number of residues, i.e., Glu438, Trp209, Val208, Arg29, Leu35, Trp346, Asn347, Asn348, and Lys206, were found to bind through hydrogen bonds of its H atom with the lone pair of oxygen of the compound, along with three ionic interactions which further increase the binding affinity of receptor-ligand. Whereas in Figure 5B (ii) the standard ligand, *i.e.,* acarbose, also showed several interactions, and all are common with compound**26** interactions and making some new interactions as well.

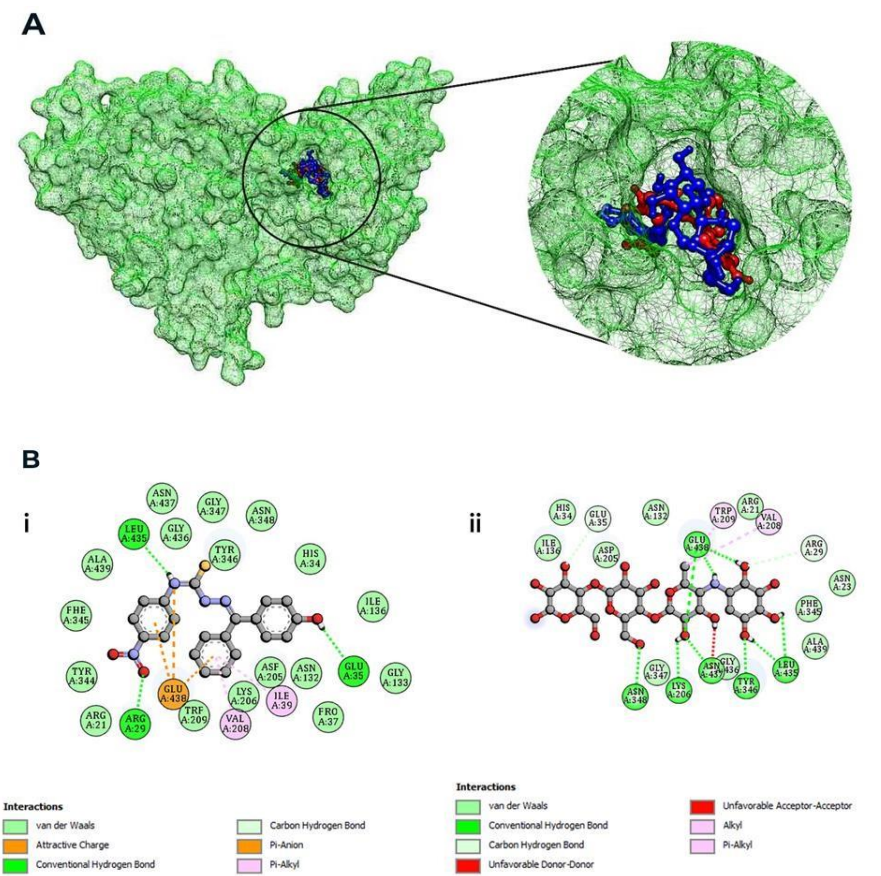

**Figure 6A:** The mesh surface protein depicts the same binding pockets of both compound **26** and acarbose, respectively. **B:** Similarly (i) and (ii) shows the interacting residues and their interaction with α-Glucosidase, respectively.

**Binding Interactions of Prolyl endopeptidase.** The above strategy was used to explore the binding approaches of compound **26** in the active site of PEP. The standard ligand Z_Pro_Prolinal of PEP was also docked. Compound **26** was found to be well accommodated in the active site of PEP, as shown in Figure 5A. The active site residue Trp597, Phe193, Phe479, and Tyr252 established a strong π–π stacking between their aromatic rings Figure 6B (i) and (ii). Similar stacking interactions have been reported earlier to protect drug structure, help in maintaining its functional effects, and facilitating its release.[57] However, Ser556, Tyr597, and Tyr601 were also mediating hydrogen bonding with the OH group of ligands, helping increase the binding affinity of the ligand with their receptor protein, as depicted in Figure S2 of the supporting information.

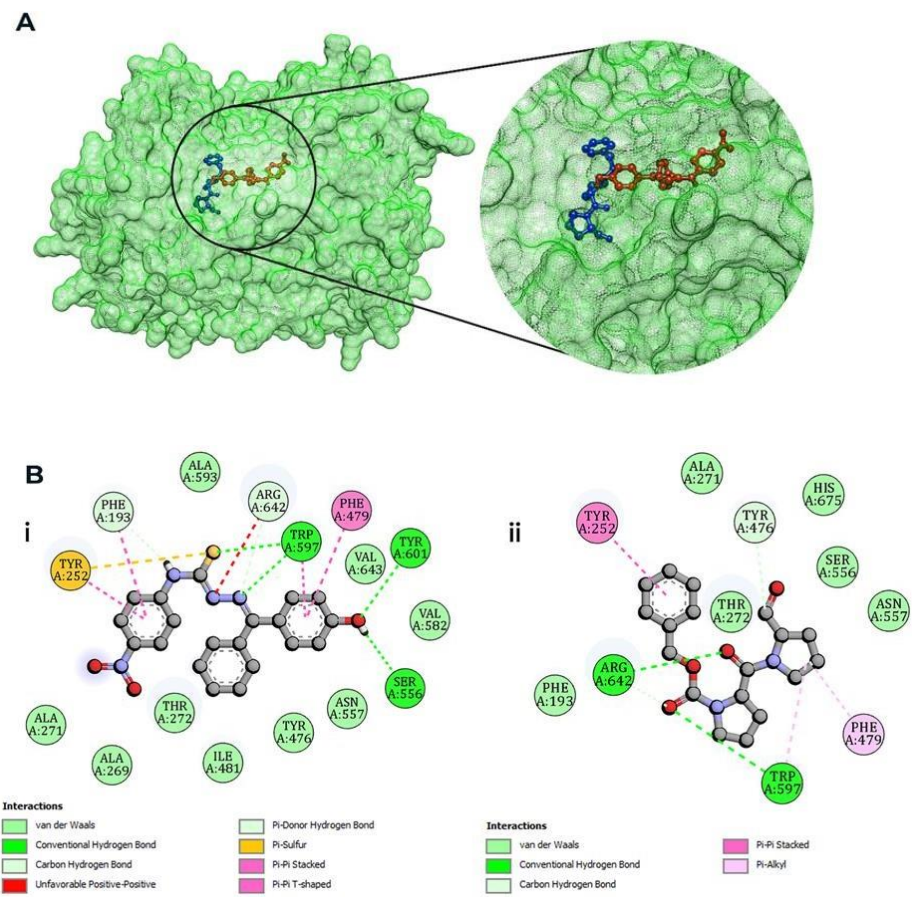

**Figure 7A:** The mesh surface protein depicts the same binding pockets of both compounds **26** and Z_Pro_Prolinal, respectively, **B**: Similarly (i) and (ii) show the interacting residues, and their interactions with prolyl endopeptidase, respectively.

The prolinal also showed similar interactions that validated our protocol as well as the potential new inhibitor *i.e.,* compound **26**. All the results were analyzed, and visualized using MGL tool, ezCADD, and UCSF Chimera. [59-61]

**Binding Free Energies.** The estimated binding free energy of the top-ranked conformation of compound 26 with $\alpha$-glucosidase is -8.25 kcal/mol. In contrast, acarbose having a binding affinity of -7.89 kcal/mol showed similarity in terms of their binding affinities and specificities. In addition, prolyl endopeptidase with compound **26** contains -7.17 kcal/mol, while the prolinal has the binding free energy of -7.63 kcal/mol, as shown in Figure 7. The binding free energies of both compounds against PEP also showed good agreement between the receptor molecules. Hence it further ensured the compound **26** as a potential drug candidate**.**

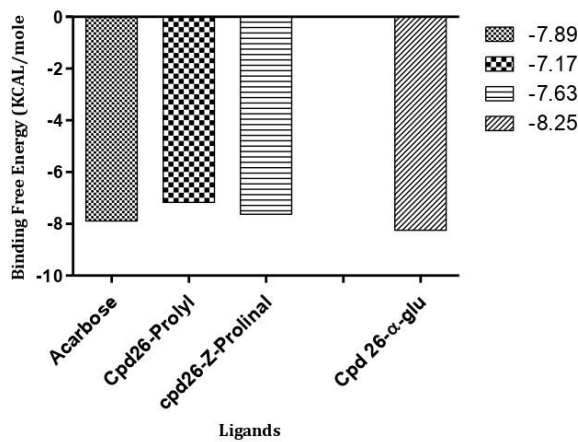

**Figure 7. The estimated binding free energies between the ligand and receptor complex**. Compound **26**, depicted here as potential inhibitor for both enzymes (α-glucosidase and prolyl endopeptidase) while their standards acarbose and prolinal, were also docked for the validation of our study.

**Statistical Analyses.** EZ-Fit (Perrella Scientific Inc., Amherst, USA) used for $IC_{50}$ calculation, and % inhibition was calculated by using the following formula: % Inhibition =100-OD of evaluated compounds * 100: Kinetics studies were done with varying concentration of substrate 0.1, 0.2, 0.4, and 0.8 mM,and *Ki* values were calculated using Grafit 7 software.

## Accession IDs
A0A0A1E2X6: A0A0A1E2X6
PPCE: Q06903

- ## ASSOCIATED CONTENT
### Supporting information

The supporting information available free of charge at

- ## AUTHOR INFORMATION
### Corresponding Author

**\*Qurat-ul-Ain**-Dr. Panjwani Center for Molecular Medicine and Drug Research, InternationalCenter for Chemical and Biological Sciences, University of Karachi, Karachi-75270, Pakistan; Orcid.org/0000-0003-4092-6900; quratulain@iccs.edu

**\*M. Iqbal Choudhary**-H. E. J. Research Institute of Chemistry, International Center for Chemical

and Biological Sciences, University of Karachi, Karachi-75270, Pakistan;Orcid.org/0000-0001-5356-3585; iqbal.choudhary@iccs.edu

**Funding** This work was supported by the HEC grant

**Notes** The authors declare no competing financial interest.

- ## ACKNOWLEDGEMENT

We thank Molecular Bank Dr. Panjwani Center, ICCBS to provide studied compounds.

- **ABBREVIATIONS**

    PEP, Prolyl endopeptidase; AD, Alzheimer's disease; ROS,reactive oxygen species; BS,benzophenone semicarbazone; DMSO, Dimethyl sulfoxide; DPPH, 2,2-diphenyl-1-picrylhydrazyl

- **REFERENCES**